\documentclass[twocolumn]{svjour3}  % ,final
\usepackage{silence} 
\usepackage{graphicx}
\usepackage{bm}                      % bold math
\usepackage{xcolor}                  % color
\usepackage{amssymb,amsmath}
\usepackage{dcolumn}                 % Align table columns on decimal point
\usepackage{multirow}
\usepackage{booktabs}
\usepackage{soul}
\usepackage[numbers,square,sort&compress]{natbib}
\usepackage[
breaklinks,pdfstartview=FitH,CJKbookmarks=true,
bookmarksnumbered=true,bookmarksopen=true,pdfborder={0 0 1},
colorlinks=true,linkcolor=blue,urlcolor=blue,anchorcolor=blue,citecolor=blue]
{hyperref}
\journalname{EPJ}

\begin{document}

\def\bea{\begin{eqnarray}} \def\eea{\end{eqnarray}}
\def\be{\begin{equation}} \def\ee{\end{equation}}
\def\bal#1\eal{\begin{align}#1\end{align}}
\def\bse#1\ese{\begin{subequations}#1\end{subequations}}
\def\rra{\right\rangle} \def\lla{\left\langle}
\def\rv{\bm{r}} \def\tv{\bm{\tau}} \def\sv{\bm{\sigma}}
\def\al{\alpha}
\def\eps{\varepsilon}
\def\ms{\,M_\odot}
\def\mmax{M_\text{max}}
\def\fm3{\;\text{fm}^{-3}}
\def\km{\;\text{km}}
\def\mev{\;\text{MeV}}
\def\fkm{f_K^\text{max}}
\def\mkm{M_K^\text{max}}
\def\mkm{M_\text{max}^{(K)}}
\def\rkm{R_\text{min}^{(K)}}
\long\def\hm#1{\color{blue}#1\color{black}}
\long\def\hj#1{\color{red}#1\color{black}}
\long\def\OFF#1{}

\title{
Properties of rapidly rotating hot neutron stars within Brueckner theory}

\OFF{
\author{
Hong-Ming Liu \inst{1}
\and
Bo-Xiu Zhou \inst{2}
\and
Zeng-Hua Li \inst{2,}\thanks{e-mail: zhli09@fudan.edu.cn (corresponding author)}
\and
Jin-Biao Wei \inst{3}
\and
H.-J. Schulze \inst{4}
}

\institute{
The Research Center for Theoretical Physics, Science School,
Qingdao University of Technology, Qingdao 266033, People's Republic of China
\and
Institute of Modern Physics, Key Laboratory of Nuclear Physics
and Ion-Beam Application, MOE, Fudan University, Shanghai, 200433, China
\and
School of Mathematics and Physics, University of Geosciences,
Wuhan 430074, People's Republic of China
\and
INFN, Sezione di Catania, Dipartimento di Fisica, Universit\'a di Catania,
Via Santa Sofia 64, 95123 Catania, Italy}
}%OFF

\author{
Hong-Ming Liu
\and
Bo-Xiu Zhou
\and
Zeng-Hua Li
\and
Jin-Biao Wei
\and
\hbox{H.-J. Schulze}
}

\institute{
H.-M. Liu \at
The Research Center for Theoretical Physics, Science School,
Qingdao University of Technology, Qingdao 266033, People's Republic of China
\and
B.-X. Zhou \at
Institute of Modern Physics, Key Laboratory of Nuclear Physics
and Ion-Beam Application, MOE, Fudan University, Shanghai, 200433, China
\and
Z.-H. Li \at
Institute of Modern Physics, Key Laboratory of Nuclear Physics
and Ion-Beam Application, MOE, Fudan University, Shanghai, 200433, China\\
e-mail: zhli09@fudan.edu.cn (corresponding author)
\and
J.-B. Wei \at
School of Mathematics and Physics, University of Geosciences,
Wuhan 430074, People's Republic of China
\and
H.-J. Schulze \at
INFN, Sezione di Catania, Dipartimento di Fisica, Universit\'a di Catania,
Via Santa Sofia 64, 95123 Catania, Italy
}

\date{Received: date / Revised version: date}

\maketitle

\abstract{
We study the properties of hot rapidly rotating neutron stars
using various equations of state based on the Brueckner-Hartree-Fock approach
at finite temperature.
The temperature dependence of
%stellar-matter properties such as the
gravitational mass, equatorial radius, Kepler frequency, moment of inertia
are analysed in detail
and universal relations are examined.
\keywords{Neutron stars \and Nuclear matter}
\OFF{
\PACS{
{21.65.-f,}  % Nuclear matter
 97.60.Jd  % Neutron stars
% 26.60.-c,  % Nuclear matter aspects of neutron stars
% 21.65.Mn,  % Equations of state of nuclear matter
% 26.60.Kp,  % Equations of state of neutron star matter
% 97.10.Cv,  % Stellar structure, interiors, evolution, nucleosynthesis, ages
% 24.10.Cn,  % Many-body theory
% 21.65.Cd,  % Asymmetric matter, neutron matter
% 13.75.Cs,  % Nucleon-nucleon interactions
% 13.75.Ev,  % Hyperon-nucleon interactions
% 26.60.Dd,  % Neutron star core
% 26.60.Gj,  % Neutron star crust
% 26.50.+x,  % Nucl. physics aspects of (super)novae and explosive environments
% 97.60.Gb,  % Pulsars
% 97.60.Bw,  % Supernovae
% 97.21.+a,  % Protostars
% 12.38.Mh,  % Quark-gluon plasma in quantum chromodynamics
% 25.75.Nq,  % Quark deconfinement, QGP production, phase transitions in RHIC
}%end PACS
}%OFF
}%end of abstract

%-------------------------------------------------------------------------------
\section{Introduction}
\label{intro}

Neutron stars (NSs),
formed as the compact remnants after core-collapse supernovae (CCSN) explosions,
offer an opportunity to probe the properties of dense matter
under extreme conditions,
as they harbour the highest binding energy per nucleon,
the strongest magnetic fields,
and the strongest spacetime curvatures in the Universe
\cite{Liebendorfer05}.
These remnants, known as proto-NSs (PNSs),
are initially extremely hot and lepton-rich,
with temperatures of the order of $\sim50\mev$.
Over a timescale of tens of seconds,
they evolve through distinct stages:
the initial neutrino-trapping phase with a high lepton fraction
($Y_l\sim0.4$), followed by a deleptonization phase
where neutrinos diffuse out and the star further heats up
%eventually settling into cold, catalyzed NSs
\cite{Burrows86,Prakash97,Pons99,Camelio17}.
When the entropy per baryon in the center of the PNS reaches its maximum,
about $2k_B$,
the deleptonization stage is finished.
An overall cooling stage follows,
with the temperature in the star steadily decreasing.
According to simulations of the CCSN dynamics,
the entropy per baryon in the PNS is approximately constant (isentropic)
during the cooling stage \cite{Thompson02,Burrows12,Kumar24}.
In this work we focus on this period,
starting about 10~s after core collapse.
Due to the strong gravitational binding,
the remnants in the early stages of evolution may exhibit high spin frequency,
limited only by mass shedding at the Kepler frequency $f_K$,
and possess higher metastable masses than the nonrotating stars.
Rotating NSs might be observed as pulsars.

\OFF{
Immediately after the core bounce of a massive star,
the core shrinks to less than $20\km$, forming a PNS,
which then starts the Kelvin-Helmholtz evolution
in the following several tens of seconds.
During this epoch,
neutrino trapping occurs because the neutrino mean free path
is much smaller than the star's radius \cite{Pons99}.
Neutrinos thus contribute to the pressure and energy density in the star,
and influence the structure and evolution of PNSs
\cite{Prakash97,Logoteta22}.
The Kelvin-Helmholtz epoch usually consists of two major evolutionary stages
\cite{Burrows86,Prakash97,Pons99,Camelio17}:
the deleptonization stage and the cooling stage.
At the beginning of the deleptonization stage,
the PNS has a high-entropy mantle and a cooler core.
Then the excess trapped electron neutrinos diffuse out of the PNS,
heating its central region,
while the net lepton and proton fractions decrease.
When the entropy per baryon in the center of the PNS reaches its maximum,
about $2k_B$,
the deleptonization stage is finished.
An overall cooling stage follows,
with the entropy as well as lepton fractions in the star steadily decreasing.
According to simulations of the supernova collapse dynamics,
the entropy per baryon in the PNS is approximately constant (isentropic)
during the cooling stage \cite{Thompson02,Burrows12,Kumar24}.
In this work we focus on this period, $\sim 5-20\;$s,
until neutrino transparency \cite{Burrows86}.
}

From the observational point of view,
currently over 3000 pulsars have been observed,
including several massive ones,
in particular the most massive
PSR J0952-0607 ($M=2.35\pm0.11\ms$) \cite{Romani25},
which puts a quite stringent lower limit on $\mmax\gtrsim2.2\ms$
of cold NSs.
The Neutron Star Interior Composition Explorer (NICER) space mission
provided simultaneous mass and radius measurements for
several pulsars
that are summarized in Table~\ref{t:psr}.
The currently fastest known pulsars
are PSR J1748-2446ad in the globular cluster Terzan 5
with spin frequency $f=716\,$Hz
and the X-ray transient XTE J1739-285 with $f=1122\,$Hz,
which are still far below the theoretical limit \cite{Hessels06}.

\begin{table}[t]%....................................................................
\centering
\renewcommand{\arraystretch}{1.3}
\setlength{\tabcolsep}{3pt}
\caption{Observational constraints on NS masses and radii.}
\footnotesize
\begin{tabular}{lccc}
%\toprule
Source & \( M / M_\odot \) & \( R \, [\text{km}] \) & Refs. \\
\midrule
J1614-2230   & \(1.908\pm0.016\)    &          & \cite{Arzoumanian18} \\
J0348+0432   & \(2.01\pm0.04\)      &          & \cite{Antoniadis13} \\
J0348+0432   & \(1.806\pm0.037\)    &          & \cite{Saffer25} \\
J0740+6620   & \(2.08\pm0.07\)      &          & \cite{Fonseca21} \\
J0952-0607   & \(2.35\pm0.11\)      &          & \cite{Romani25} \\
\midrule
\multirow{5}{*}{J0030+0451}
& \(1.44^{+0.15}_{-0.14}\) & \(13.02^{+1.24}_{-1.06}\) & \cite{Miller19,Riley19} \\
& \(1.36^{+0.15}_{-0.16}\) & \(12.71^{+1.14}_{-1.19}\) & \cite{Riley21,Miller21} \\
& \(1.40^{+0.13}_{-0.12}\) & \(11.71^{+0.88}_{-0.83}\) & \cite{Vinciguerra24} \\
& \(1.70^{+0.18}_{-0.19}\) & \(14.44^{+0.88}_{-1.05}\) & \cite{Vinciguerra24} \\
& \(1.43^{+0.20}_{-0.17}\) & \(12.68^{+1.31}_{-1.04}\) & \cite{Kini26} \\
\midrule
\multirow{4}{*}{J0740+6620}
& \(2.072\pm0.067\)    & \(12.39^{+1.30}_{-0.98}\) & \cite{Riley21} \\
& \(2.08\pm0.07\)      & \(13.7^{+2.6}_{-1.5}\)    & \cite{Miller21} \\
& \(2.073\pm0.069\)    & \(12.49^{+1.28}_{-0.88}\) & \cite{Salmi24} \\
& \(2.08\pm0.07\)      & \(12.92^{+2.09}_{-1.13}\) & \cite{Dittmann24} \\
\midrule
\multirow{2}{*}{J0437-4715}
& \(1.418\pm0.037\)    & \(11.36^{+0.95}_{-0.63}\) & \cite{Choudhury24} \\
& \(1.418\pm0.044\)    & \(13.7\pm1.8\)             & \cite{Miller26} \\
\midrule
J0614-3329 & \(1.44^{+0.06}_{-0.07}\) & \(10.29^{+1.01}_{-0.86}\) & \cite{Mauviard25} \\
\midrule
J1231-1411 & \(1.04^{+0.05}_{-0.03}\) & \(12.6\pm0.3\ \text{or}\ 13.5^{+0.3}_{-0.5}\) & \cite{Salmi24b} \\
\midrule
\multirow{7}{*}{$\displaystyle\text{GW170817}\atop\displaystyle\text{+NICER}$}
& \(2.08\) & \(12.35^{+0.75}_{-0.75}\) & \cite{Miller21} \\
& \(2.0\)  & \(12.33^{+0.70}_{-1.34}\) \text{or} \(11.55^{+0.94}_{-1.09}\) & \cite{Rutherford24} \\
& \(2.0\)  & \(11.99^{+0.85}_{-1.25}\) \text{or} \(11.20^{+1.05}_{-0.94}\) & \cite{Mauviard25} \\
& \(1.4\)  & \(12.28^{+0.50}_{-0.76}\) \text{or} \(12.01^{+0.56}_{-0.75}\) & \cite{Rutherford24} \\
& \(1.4\)  & \(12.05^{+0.56}_{-0.79}\) \text{or} \(11.71^{+0.71}_{-0.63}\) & \cite{Mauviard25} \\
& \(1.4\)  & \(12.57^{+0.49}_{-0.48}\) & \cite{Dittmann24} \\
& \(1.4\)  & \(12.56^{+0.47}_{-0.47}\) & \cite{Miller26} \\
\end{tabular}
\label{t:psr}
\end{table}%....................................................................

The gravitational wave (GW) signal of the binary neutron star (BNS)
merger GW170817 \cite{Abbott17,Abbott18}
accompanied by an electromagnetic counterpart AT2017gfo
spanning the X-ray to radio bands,
as well as a short $\gamma$-ray burst GRB170817A,
offer a new avenue to probe the behavior of matter at high densities,
high temperatures and high spin frequencies.
The analysis of tidal deformability gave an estimate of
$\Lambda_{1.4}$ = 190$^{+390}_{-120}$ \cite{Abbott18}.
Combining GW observations from GW170817 with the results of NICER
\cite{Miller19,Riley19,Miller21,Riley21,Pang21,Raaijmakers21,%
Salmi24,Dittmann24,Choudhury24,Rutherford24,Mauviard25,Miller26},
the maximum mass for slowly- or non-rotating NSs
is likely in the range of $\mmax=2.15-2.30\ms$
\cite{Margalit17,Shibata17,Rezzolla18,Shibata19,%
Shao20,Nathanail21,Rutherford24,Mauviard25,Miller26}.

While astrophysical constraints regarding hot rotating NSs
are still not available,
the study of thermal effects is of great significance for the simulation
of the dynamical evolution of CCSNs \cite{Oconnor18,Burrows19},
proto-NSs \cite{Prakash97,Nicotra06,Burgio11}
and BNS mergers \cite{Baiotti17,Ruiz20}.
It also probes the influence of temperature
on the specific properties of rotating NSs,
such as gravitational mass, equatorial radius, Kepler frequency,
moment of inertia (MOI) and others.
For this purpose, publicly available finite-temperature
equations of state (EOSs) are essential \cite{Oertel17,Burgio21}.
Among those, we mention the EOSs
Shen11 \cite{Shen98,Shen11},
Shen20 \cite{Shen20},
HS(DD2) \cite{Hempel10,Typel10},
SFHx \cite{Steiner13} and
FSU2H \cite{Tolos16}
based on the covariant density-functional theory,
the Lattimer-Swesty EOS \cite{Lattimer91}
based on nonrelativistic Skyrme interactions,
the TNTYST \cite{Togashi17},
SRO (APR) \cite{Schneider19}
and the Brueckner-Hartree-Fock (BHF) EOS V18
\cite{Li08a,Li08,Burgio10,Lu19,Wei21,Liu22}
based on microscopic theoretical calculations.
Unlike non-rotating NSs,
rapidly rotating NSs are typically modeled as axisymmetric rigid bodies
and solved numerically in GR.
Over the past few decades,
several groups have developed numerical methods for simulating
axisymmetric rotating stellar structures
\cite{Komatsu89,Komatsu89b,Weber91,Weber91b,Cook92,Cook94,Salgado94,%
Stergioulas95}.

Over recent years, considerable effort has been devoted to
understanding how thermal effects affect the integral properties of
(non-)rotating NSs
\cite{Batra18,Soma20,Wei21,Koliogiannis21,Farrell24,Tsiopelas26,
Yasin2020,Lu19,Figura20,Figura21,Constantinou2015,Nilaksha25,Barman2025}.
For example, \cite{Wei21} systematically compared the
thermodynamic properties of nuclear matter using various finite-temperature EOSs
and investigated the influence of thermal effects
on the maximum mass of non-rotating NSs.
\cite{Soma20} studied the properties of hot,
rapidly rotating merger remnants at fixed entropy per baryon,
showing that thermal pressure enlarges the stellar radius
and lowers the Kepler frequency compared to cold configurations.
\cite{Tsiopelas26} investigated how variations
in the symmetry energy slope $L_\text{sym}$
affect the global properties of isentropic stars,
concluding that lower $L_{\rm sym}$ leads to higher maximum masses
and smaller radii.
\cite{Yasin2020} showed that the nucleon effective mass 
determines thermal nucleonic contributions to the EOS, 
governing PNS contraction and supernova shock evolution.
\cite{Nilaksha25} constructed finite-temperature EOSs 
within a non-linear relativistic mean field model, 
and investigated thermal effects on 
NS macroscopic properties and $f$-mode oscillations.

Parallel investigations have focused on the (EOS-independent)
universal relations (URs) of (non-)rotating NSs at finite temperature
\cite{Doneva13,Martinon14,Chakrabarti14,Marques17,Lenka19,Raduta20,
Khadkikar21,Patkos22,Largani22,Cecilia2015,Alejandro2019,
Kumar2025,Ghosh26,Christian2020}.
For example,
early works on PNS evolution reported significant violations
of the $I$-Love-$Q$ relations during the hot, lepton-rich phase,
with universality being restored only as the star cools and deleptonizes
\cite{Doneva13,Martinon14,Marques17,Lenka19,Chakrabarti14}.
Subsequent work demonstrated that these relations hold with good accuracy
when thermodynamic conditions are kept fixed \cite{Raduta20}.
In particular, URs linking the maximum masses of Keplerian and
static configurations have been established for hot stars
and applied to infer constraints on the cold NS mass from the GW170817
\cite{Khadkikar21}.
Moreover, 
URs connecting $f$-mode frequencies with stellar compactness, 
first established for cold NSs \cite{Cecilia2015}, 
have been extended to PNSs at finite temperature \cite{Alejandro2019,Kumar2025,Ghosh26} 
and to rapidly rotating configurations \cite{Christian2020}.

Theoretical studies indicate that the maximum mass of a rapidly rotating NS
can increase by approximately 20\% compared to its static counterpart
\cite{Cook94,Cook94b,Bonazzola98},
which offers a potential explanation for the massive secondary object
observed in the GW190814 event \cite{Abbott20}
with a mass of 2.59$^{+0.08}_{-0.09}\ms$
as a rapidly rotating NS
\cite{Zhang20,Astashenok20,Dexheimer21}.
Some other studies also pointed out that the light companion in GW190814
might be a slowly rotating or non-rotating NS or a black hole,
which are sensitive to the stiffness of the selected EOSs at high densities
\cite{Huang20,Tsokaros20,Fattoyev20,Tews21}.

In our previous works \cite{Lu19,Liu22}
we have constructed the finite-temperature EOS
based on the BHF theory including three-body forces
to simulate the dynamical evolutions of BNS mergers
\cite{Figura20,Figura21}
and investigate the properties of hot NSs.
This work aims to extend our study to rotating NSs
and examine the effects of thermal conditions on parameters
such as gravitational mass, equatorial radius, Kepler frequency
and related properties.
For modeling hot NSs,
isentropic models with fixed lepton fraction are widely adopted
and considered suitable to describe the early stage of PNS evolution.
However,
the present work does not aim to model the detailed evolution of PNSs,
but to focus on the pure thermal effects on rotating PNSs,
for which neutrino-free isentropic and isothermal approximations are sufficient.

Our paper is organized as follows.
The construction of the EOS of hot rotating NSs is briefly reviewed
in Sec.~\ref{s:form}.
Their properties, such as the gravitational mass,
equatorial radius, Kepler frequency
and related quantities
are discussed in Sec.~\ref{s:res}.
Conclusions are drawn in Sec.~\ref{s:conc}.

%-------------------------------------------------------------------------------
\section{Formalism}
\label{s:form}

%-------------------------------------------------------------------------------
\subsection{Brueckner-Bethe-Goldstone theory at finite temperature}

In hot stellar matter, the free-energy density is given as
\be
 f = f_N + f_L \:,
\ee
where $f_N$ and $f_L$ represent the contributions of nucleons and leptons
($e$, $\mu$, $\nu_e$, $\nu_\mu$ and their antiparticles),
respectively.
In this work, we employ the BHF approach for asymmetric nuclear matter
at finite temperature
\cite{Bloch58,Bloch59a,Bloch59b,Lejeune86,Baldo99,Burgio10}
to calculate the nucleonic contribution.
In this approach the essential ingredient is the interaction matrix $K$, which,
along with the single-particle (s.p.) energy $e_1$ and potential $U_1$,
satisfies the self-consistent equations
\bal
 K(\rho,x_p;E) &= V +
 \text{Re}\sum_{1,2} V \frac{|12\rangle \bar{n}_1 \bar{n}_2 \langle12|}
 {E - e_1 - e_2 + i0} K(\rho,x_p;E) \:,
\label{e:k}
\\
 e_1 &= \frac{k_1^2}{2m_1} + U_1 \:,
\label{e:e}
\\
 U_1(\rho,x_p) &= \sum_2 n_2 \langle12| K(\rho,x_p;e_1+e_2) |12\rangle_a \:,
\label{e:u}
\eal
where
$\bar{n}_i \equiv 1-n_i$,
$E$ is the starting energy, and
$x_p$ = $\rho_p$/$\rho$ is the proton fraction with $\rho_p$ and $\rho$
being the proton and total baryon densities, respectively.
The multi-indices 1,2 generally denote the momentum, isospin and spin.
$K$ and $V$ are understood as operators in this space. 
At finite temperature,
\be
 n_i(k) = \left[ \exp\left(\frac{e_i(k)-\tilde\mu_i}{T}\right) + 1 \right]^{-1}
\ee
is a Fermi distribution,
where $\tilde\mu_i$ is an auxiliary chemical potential.
For a given density and temperature,
Eqs.~(\ref{e:k}-\ref{e:u})
have to be solved self-consistently along with the condition
\be
 \rho_i = 2 \sum_k n_i(k) \:,
\ee
until the $K$ matrix reaches convergence.

In previous works \cite{Baldo99,Nicotra06},
we introduced the so-called frozen-correlations approximation,
in which the s.p.~potential $U_i(k)$ for the component $i$ at $T \neq 0$
is approximated by the one calculated at $T = 0$.
Within this approximation,
the nucleonic free energy density $f_N$ and entropy density $s_i$
are given by the following simplified expressions:
\bal
 f_N &= \sum_{i=n,p} \left[ 2\sum_k n_i(k)
 \left( \frac{k^2}{2m_i} + \frac12 U_i(k) \right) - T s_i \right] \:,
\\
 s_i &= -2\sum_k \Big[n_i(k) \ln n_i(k) + \bar{n}_i(k) \ln\bar{n}_i(k)\Big] \:,
\eal
and all other thermodynamic quantities can be derived in a consistent way,
namely, the ``true'' chemical potentials $\mu_i$, pressure $p$,
entropy density $s$
and internal energy density $\eps$ are defined as
\bal
 \mu_i &= \frac{\partial f}{\partial\rho_i} \:,
\label{e:mu}
\\
 p &= \rho^2\frac{\partial(f/\rho)}{\partial\rho} = \sum_i \mu_i \rho_i - f \:,
\label{e:p}
\\
 \eps &= f + Ts\ ,\ \ s = - \frac{\partial f}{\partial T} \:.
\label{e:eps}
\eal

\begin{figure*}[t]%.............................................................
\vspace{1mm}\hspace{-1mm}
\includegraphics[width=1\textwidth]{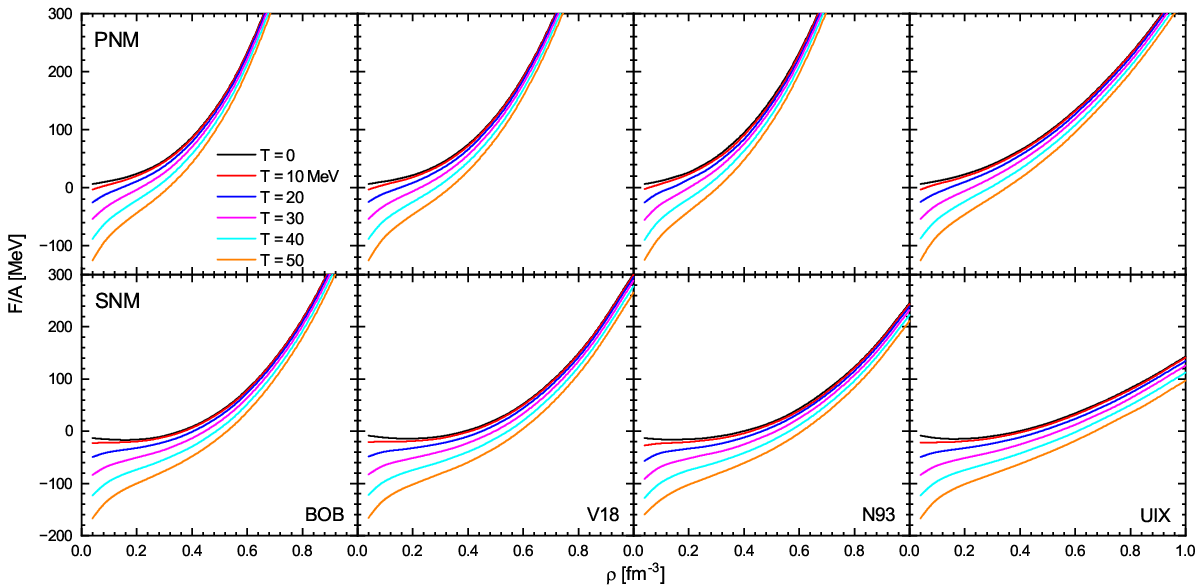}
\vspace{-5mm}
\caption{
Free energy per nucleon as a function of nucleon density
for symmetric (lower panels) and pure neutron (upper panels) matter
with different EOSs
for temperatures between 0 and 50 MeV in steps of 10 MeV.}
\label{f:fa}
\end{figure*}%..................................................................

\begin{table}[b]%...............................................................
\caption{
Parameters of the EOS fits for the free energy per nucleon $F/A$,
Eq.~(\ref{e:fit}),
for symmetric nuclear matter (SNM) and pure neutron matter (PNM).}
\small
\renewcommand{\arraystretch}{1.2}
\setlength{\tabcolsep}{3pt}
\begin{tabular}{lcrrrrrrrrr}
\toprule
 &  & $a$ & $b$ & $c$ & $d$
 & $\tilde{a}$ & $\tilde{b}$ & $\tilde{c}$ & $\tilde{d}$ & $\tilde{e}$ \\
\midrule
BOB & SNM & -63 & 499 & 2.69 & -9 & -123 & 204 & -59 &  74 & 2.29 \\
    & PNM &  54 & 857 & 2.90 &  4 &  -84 & 153 & -60 &  69 & 2.21 \\
V18 & SNM & -61 & 370 & 2.66 & -7 & -146 & 208 & -63 &  79 & 2.28 \\
    & PNM &  41 & 665 & 2.79 &  5 &  -91 & 155 &-104 & 112 & 2.12 \\
N93 & SNM & -42 & 298 & 2.61 &-12 & -142 & 211 & -64 &  87 & 2.35 \\
    & PNM &  67 & 743 & 2.71 &  4 &  -95 & 154 & -35 &  46 & 2.44 \\
UIX & SNM &-213 & 358 & 1.53 & -3 & -184 & 198 &-331 & 344 & 2.05 \\
    & PNM &  29 & 323 & 2.12 &  6 & -118 & 154 &-106 & 114 & 2.12 \\
\bottomrule
\end{tabular}
\label{t:f}
\end{table}%....................................................................

The realistic nucleon-nucleon ($NN$) interactions $V$
adopted in the present calculations include
the Bonn B (BOB) \cite{Machleidt1987}
or the Nijmegen 93 (N93) \cite{Nagels78,Stoks94}
potential supplemented with a microscopic three-body force (TBF)
employing the same meson-exchange parameters as the two-body potentials,
and the Argonne V18 potential \cite{Wiringa95},
combined either with a microscopic \cite{Grange89,Li08a}
or a phenomenological \cite{Carlson83,Zhou04} TBF,
which are labeled as V18 and UIX, respectively.
The TBF is reduced to an effective, density-dependent, two-body force
by averaging over the third nucleon in the medium in the BHF calculation,
\be
 \overline{V}_{12}(r_{12}) = \rho \int\! d^3\rv_3 \sum_{\sigma_3,\tau_3}
 g_{13}(r_{13})^2 g_{23}(r_{23})^2 V_{132} \:,
\ee
the average being weighted by the BHF correlation functions
$g_\alpha(r) = 1-\eta_\alpha(r)$,
related to the $r$-space defect functions in the partial waves $\alpha$,
\be
 \eta_\alpha(r) = j_\alpha(r) - u_\alpha(r) \:,
\ee
where $j_\alpha$ and $u_\alpha$ are the free wave function (Bessel function)
and the correlated wave function
representing the nucleon-nucleon in-medium correlations, respectively
\cite{Li08a,Grange89,Zuo02,Zhou04,Li16}.
This produces an effective two-nucleon potential with operator structure
\bal
 \overline{V}_{\!12}(\rv) =&\; (\tv_1\cdot\tv_2)
 (\sv_1\cdot\sv_2) V_C(r) + (\sv_1\cdot\sv_2) V_S(r) + V_I(r)
\nonumber\\
 &+ S_{12}(\hat\rv) \left[ (\tv_1\cdot\tv_2) V_T(r) + V_Q(r) \right] \:,
\eal
where $S_{12}(\rv)=3(\sv_1\cdot\hat\rv)(\sv_2\cdot\hat\rv)-\sv_1\cdot\sv_2$
is the tensor operator
and the five components $V_C,V_S,V_I,V_T,V_Q$
depend on the nucleon densities $\rho_i$.

Fig.~\ref{f:fa} shows the nucleonic free energy per nucleon $F/A$
as function of the baryon density $\rho$
using the four BHF EOSs
for symmetric nuclear matter (SNM) ($x_p=1/2$, lower panel)
and pure neutron matter (PNM) ($x_p=0$, upper panel)
for temperatures between 0 and 50 MeV.
At $T=0$, where the free energy coincides with the internal energy,
the SNM curve is the usual nuclear-matter saturation curve,
which reveals the fundamental properties of nuclear matter
and its behavior at different densities.
One notes that at high densities
the PNM EOSs are stiffer than the SNM ones,
and that the relative temperature effects for PNM are smaller,
both due to the higher Fermi energy of the neutrons at given density.

In all cases, the following functional forms provide excellent parametrizations
of the numerical results
\cite{Lu19,Liu22}
in the required ranges of density
($0.05\fm3 \lesssim \rho \lesssim 1\fm3$) and temperature
($5\mev < T < 50\mev$):
\bal
 \frac{F}{A}(\rho,T) =&\; a\rho + b\rho^c + d
\nonumber\\
 & + \tilde{a} t^2\rho + \tilde{b} t^2 \ln(\rho)
 + ( \tilde{c}t^2 + \tilde{d} t^{\tilde{e}} )/\rho \:,
\label{e:fit}
\eal
where $t\equiv T/(100\mev)$
and $F/A$ and $\rho$ are given in MeV and fm$^{-3}$, respectively.
The parameters of the different fits are listed in Table~\ref{t:f}.
The rms deviations between the fit and data are better than 1 MeV for all EOSs.

For asymmetric matter,
it turns out that the dependence on the proton fraction
at zero temperature \cite{Burgio10,Bombaci94}
and finite temperature \cite{Li21,Liu22}
can be well approximated by a parabolic law:
\be
 \frac{F}{A}(\rho,T,x_p) \approx \frac{F}{A}(\rho,T,0.5)
 + (1-2x_p)^2 \frac{F_\text{sym}}{A}(\rho,T) \:,
\ee
where $F_\text{sym}(\rho,T)$ is the free symmetry energy:
\be
 \frac{F_\text{sym}}{A}(\rho,T) = \frac{F}{A}(\rho,T,x_p=0)
 - \frac{F}{A}(\rho,T,x_p=0.5) \:.
\label{e:fsym}
\ee
Therefore,
for the treatment of $\beta$-stable matter,
it is only necessary to provide parameterizations for SNM and PNM,
as shown in Eq.~(\ref{e:fit}).
The free symmetry energy determines the composition of stellar matter,
which is affected by both the strong and weak interactions involving leptons.
We have shown in \cite{Li21}
that going beyond the parabolic approximation has negligible effects
on the results for NS structure.

\begin{figure*}[t]%.............................................................
\vspace{0mm}
\includegraphics[width=1\textwidth]{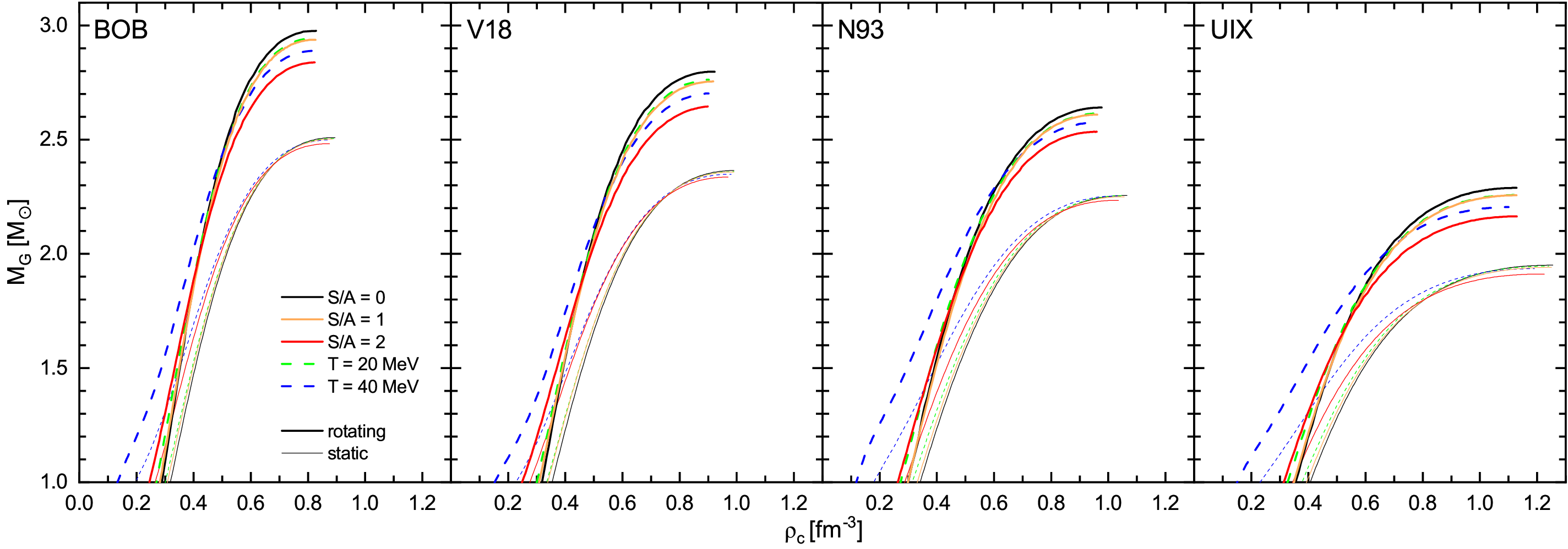}
\includegraphics[width=1\textwidth]{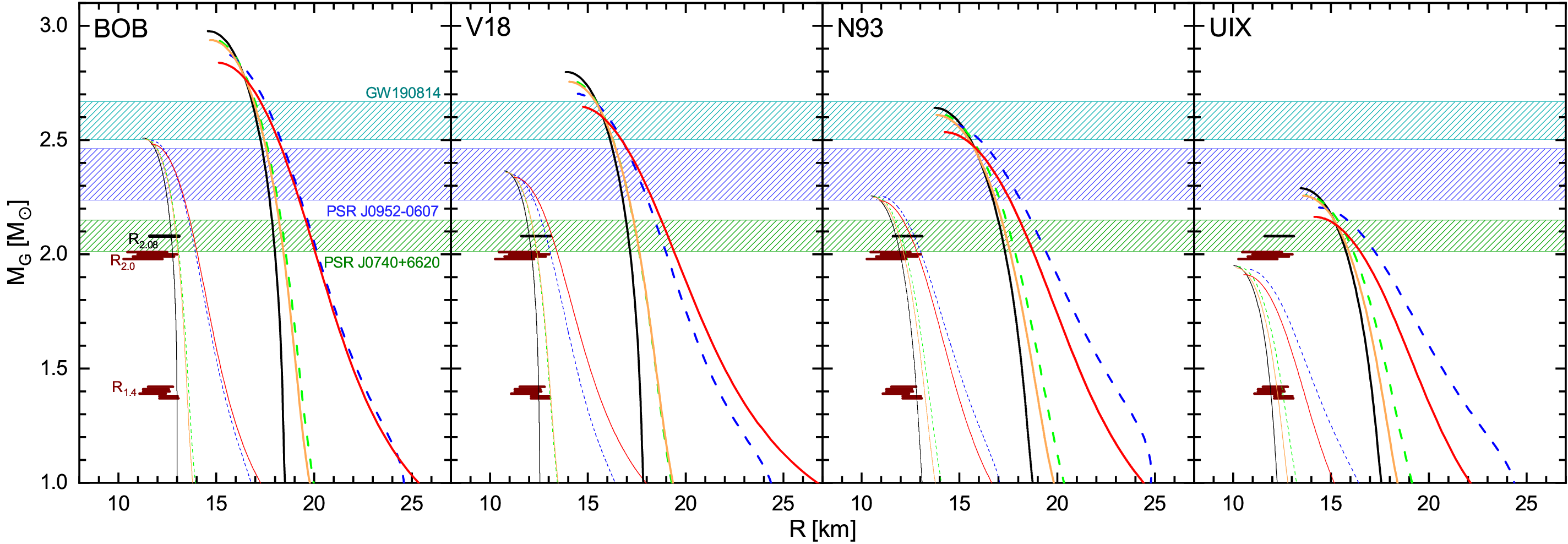}
\vspace{-4mm}
\caption{
Gravitational mass vs central density (upper panels) and radius (lower panels)
for static (thin curves) and uniformly rotating at $f_K$ (thick curves) stars
%Thick (thin) solid curves denote static (Keplerian) sequences.
and different thermal conditions:
temperatures $T=0,20,40\mev$
and constant entropy per baryon $S/A=1,2$.
All curves end at their respective $\mmax$ configuration.
The shaded areas indicate the recent constraints on mass from
PSR J0740+6620 ($M=2.08\pm0.07\ms$) \cite{Fonseca21},
PSR J0952-0607 ($M=2.35\pm0.11\ms$) \cite{Romani25} and
the GW190814 event ($M=2.50$-$2.67\ms$) \cite{Abbott20}.
%together with the constraints from the GW170817 event \cite{Abbott18}
%and the mass-radius constraints on PSR J0030+0451 and PSR J0740+6620
%from the NICER mission \cite{Riley19,Miller19,Riley21,Miller21}.
The horizontal bars indicate the constraints for the cold EOS
on $R_{2.08}$, $R_{2.0}$, and $R_{1.4}$
obtained in the combined NICER+GW170817 data analyses
of \cite{Miller21,Pang21,Raaijmakers21} (black bars)
and the recent
\cite{Rutherford24,Dittmann24,Mauviard25,Miller26} (brown bars).
}
%\hj{J0952-060 has changed ; remove  GW170817, J0030+0451, J0740+6620}
\label{f:mr}
\end{figure*}%..................................................................

%-------------------------------------------------------------------------------
\subsection{Composition and EOS of hot stellar matter}

In this paper,
we focus exclusively on $\beta$-stable and neu-\\
trino-free nuclear matter
at finite temperature.
While this may not be the most accurate approximation for merger simulations
\cite{Alford18,Figura21,Hammond21},
it is certainly inadequate for supernovae
and the initial seconds of a PNS's life,
where the presence and effects of trapped neutrinos play a crucial role
\cite{Prakash97,Pons99,Burgio11,Kaplan14,Lattimer16,Raithel2021,Rahimi23,Zheng25}.
It becomes a reasonable approximation for the later-stage evolution
of PNSs after about ten seconds,
where the temperature is still of the order of $20\mev$
\cite{Zheng25}.
We therefore focus on this temperature range.

For $\beta$-stable nuclear matter,
which includes nucleons and leptons as the relevant degrees of freedom,
the composition is determined by enforcing chemical equilibrium
and the charge-neutrality condition at a given baryon density~$\rho$,
\bal
 \mu_n - \mu_p &= \mu_e = \mu_\mu \:,
\\
 \sum_i Q_i\rho_i &= 0 \:,
\eal
where the various chemical potentials are derived
from the total free energy density $f$ according to Eq.~(\ref{e:mu}).
One can then calculate the composition of $\beta$-stable stellar matter.
Furthermore, the total pressure $p$ and the internal energy density $\eps$
can be calculated using the thermodynamic relations
Eqs.~(\ref{e:p}) and (\ref{e:eps}).

We mention that the BHF approach provides only the EOS
for the homogeneous nuclear matter region $\rho \gtrsim 0.08\fm3$.
We attach for the outer crustal part the Shen20 EOS \cite{Shen20}.
\OFF{
We attach for the outer part a cold crust,
using in the medium-density regime
($0.001\fm3 \lesssim \rho \lesssim 0.08\fm3$)
the well-known Negele-Vautherin EOS \cite{Negele73} for the inner crust,
while for the low-density regime
($\rho \lesssim 0.001 ~\text{fm}^{-3}$)
we employ the EOSs by Baym-Pethick-Sutherland \cite{Baym71}
and Feynman-Metropolis-Teller \cite{Feynman49} to describe the outer crust.
}
The merging of finite-temperature Lattimer-Swesty \cite{Lattimer91}
or Shen \cite{Shen98,Shen98b} EOSs with high-density BHF EOS
was studied in detail in \cite{Burgio10},
indicating that the maximum-mass domain is essentially unaffected
by the structure of the low-density transition region.
The choice of the crust model may slightly impact the estimations
of radius and associated deformability,
typically around 1$\%$ for $R_{1.4}$
\cite{Burgio10,Baldo14,Fortin16}.
Recent studies have shown that even completely ignoring the crust of a NS
does not significantly affect its radius and deformability
\cite{Tsang19,Perot20}.

\begin{figure*}%................................................................
\vspace{-0mm}\hspace{-1mm}
\includegraphics[width=1\textwidth]{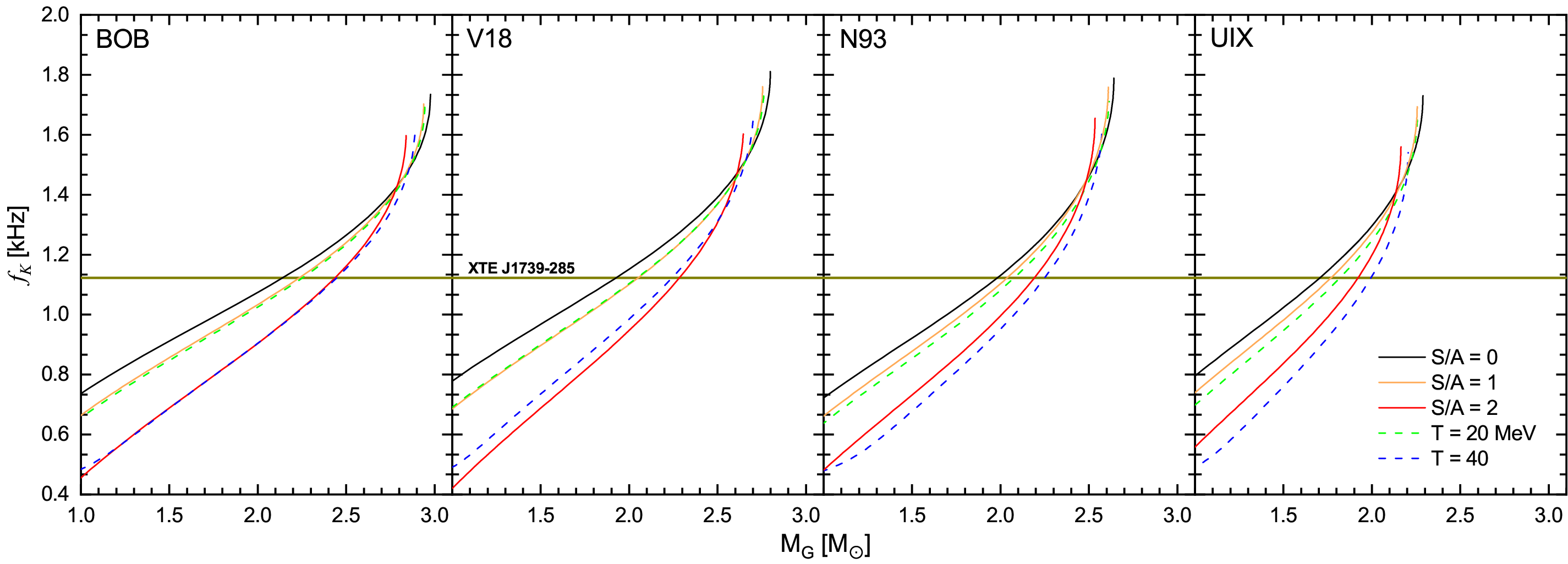}
\vspace{-4mm}
\caption{
Kepler frequency as a function of gravitational mass for different EOSs
and thermal conditions.
The horizontal line is the current observational limit 1122 Hz
from XTE J1739-285 \cite{Kaaret07}.
}
\label{f:fk}
\end{figure*}%..................................................................

%-------------------------------------------------------------------------------
\subsection{Rapidly rotating NSs}

Rotating NSs can be described by the stationary axisymmetric space-time metric
\bal
 ds^2 =& -e^{\gamma+\rho} dt^2 + e^{2\beta} (dr^2 + r^2 d\theta^2)
\nonumber\\
 &+ e^{\gamma-\rho} r^2 \sin^2\theta (d\varphi - \omega dt)^2 \:,
\eal
where $\gamma$, $\rho$, $\beta$ and $\omega$
are functions of $r$ and $\theta$ only.
Normally, the matter inside the NS is approximated by a perfect fluid
such that the energy momentum tensor can be given by
\be
 T^{\mu\nu} = (\eps + p) v^\mu v^\nu - p g^{\mu\nu} \:,
\ee
where $p$ and $\eps$ are the fluid pressure and energy density,
and $v^\mu$ and $g^{\mu\nu}$ are the four-velocity and the metric tensor,
respectively.
To solve the Einstein field equations for rotating stars,
we use the public RNS code \cite{Komatsu89,Komatsu89b,RNS}.

%-------------------------------------------------------------------------------
\section{Results}
\label{s:res}

In the following, we present the results of our numerical calculations
within the BHF theory.
We discuss the thermal effects on several properties of rapidly rotating NSs,
such as the mass and radius, Kepler frequency, MOI, Kerr parameter,
and ratio of rotational kinetic energy and gravitational potential energy.

%-------------------------------------------------------------------------------
\subsection{Mass and radius}

In Fig.~\ref{f:mr} we show the relations of gravitational mass
vs central density (upper panels) and radius (lower panels)
for static (thin curves)
and rapidly rotating NSs at the Kepler sequence (thick curves),
constructed with BOB, V18, N93, and UIX EOSs within BHF theory
for temperatures $T=0,20,40\mev$
and constant entropy per baryon $S/A=1,2$.
All sequences end at their maximum gravitational mass configurations.
The figure also shows up-to-date observational constraints on mass and radii
of the cold EOS,
as detailed in the caption.

\OFF{
The recent mass constraints for the cold EOS from
PSR J0740+6620 ($M=2.08\p0.07\ms$) \cite{Fonseca21},
PSR J0952-0607 ($M=2.35\pm0.17\ms$) \cite{Romani22}
and from the GW190814 event ($M=2.50-2.67\ms$) \cite{Abbott20},
as well as constraints on radii
from the combined NICER+GW170817 data analyses
\cite{Miller21,Pang21,Raaijmakers21,
Rutherford24,Dittmann24,Mauviard25,Miller26}
are also shown.

The mass-radius results of the NICER mission for the pulsars J0030+0451
\cite{Miller19,Riley19,Vinciguerra24},
and J0740+6620
\cite{Miller21,Riley21,Raaijmakers21,Pang21,Salmi24,Dittmann24}
%and J0437-4715 \cite{Choudhury24}
% PSR J1231-1411 \cite{Salmi24b}
are also plotted in the figure.
The combined analysis of those pulsars together with GW170817
yields improved limits on
$R_{1.4}$ \cite{Raaijmakers21,Pang21,Rutherford24},
$R_{2.0}$ \cite{Rutherford24} and
$R_{2.08}$ \cite{Miller21},
which are shown as horizontal bars.
}

As can be seen, the mass and radius of static and rotating NSs
depend quite sensitively on the EOS model.
For the cold static NSs (thin black solid curves),
the stiffest BOB EOS predicts a larger maximum mass and radius
than the other EOSs and satisfies even the mass constraint
for the heavy object of GW190814. %($M=2.50-2.67\ms$).
The maximum masses of all EOSs apart from UIX
fulfill the current observational lower limits
from PSR J0740+6620 and PSR J0952-0607.
The V18 is clearly preferred as nucleonic EOS,
also in terms of its maximum mass.
We will not delve further into the discussion of NS mass and radius here,
but refer to recent literature \cite{Wei21,Wei19,Sun21,Burgio21}
for more details.

The maximum masses for the cold static configurations
and the ones in the Kepler sequences
for BOB, V18, N93, and UIX EOSs
are $\mmax=2.53, 2.39, 2.28, 1.\\
99\ms$
and $\mkm =3.02, 2.85, 2.69, 2.35\ms$, respectively,
satisfying the EOS-independent UR
$\mkm \approx 1.2\mmax$
\cite{Cook94,Lasota96,Breu16,Khadkikar21}.

The maximum gravitational mass of static NSs
shows a weak dependence on temperature,
with a slight reduction compared to cold configurations for all EOSs,
due to the strong compensation between nucleonic and leptonic contributions
to the thermal pressure in beta-stable matter within the BHF approach
\cite{Lu19,Liu22,Figura21}.
The finite-temperature reduction of $\mmax$ is much larger
for fast-rotating NSs,
because thermal effects increase the equatorial radius of rotating NSs
and become more pronounced as the gravitational mass decreases
due to the expansion of the outer core and crust at finite temperature,
which enhances the centrifugal force at a specific rotation frequency.
As a result,
at finite temperature the Kepler limit is reached at a lower frequency,
as shown in Fig.~\ref{f:fk},
leading also to a reduction of the maximum gravitational mass $\mkm$
at that frequency.
Furthermore,
at fixed $S/A$ or $T$,
the maximum masses for the static configurations
and the ones in the Kepler sequences
also satisfy a EOS-independent linear UR
$\mkm \approx C(T) \mmax$,
where $C(T)$ decreases slightly with increasing $S/A$ or $T$
compared to the cold case \cite{Khadkikar21},
due to the above mechanism.

It is important to note again that in the current work
we assume neutrino-free conditions inside PNSs.
Neutrino effects on thermal energy density and pressure
may significantly influence (reduce) the maximum mass of (P)NSs
\cite{Nicotra06,Burgio10,Paschalidis12,Marques17,Khadkikar21}.

\begin{figure}%.................................................................
\vspace{-2mm}\hspace*{-0mm}
\includegraphics[width=0.44\textwidth]{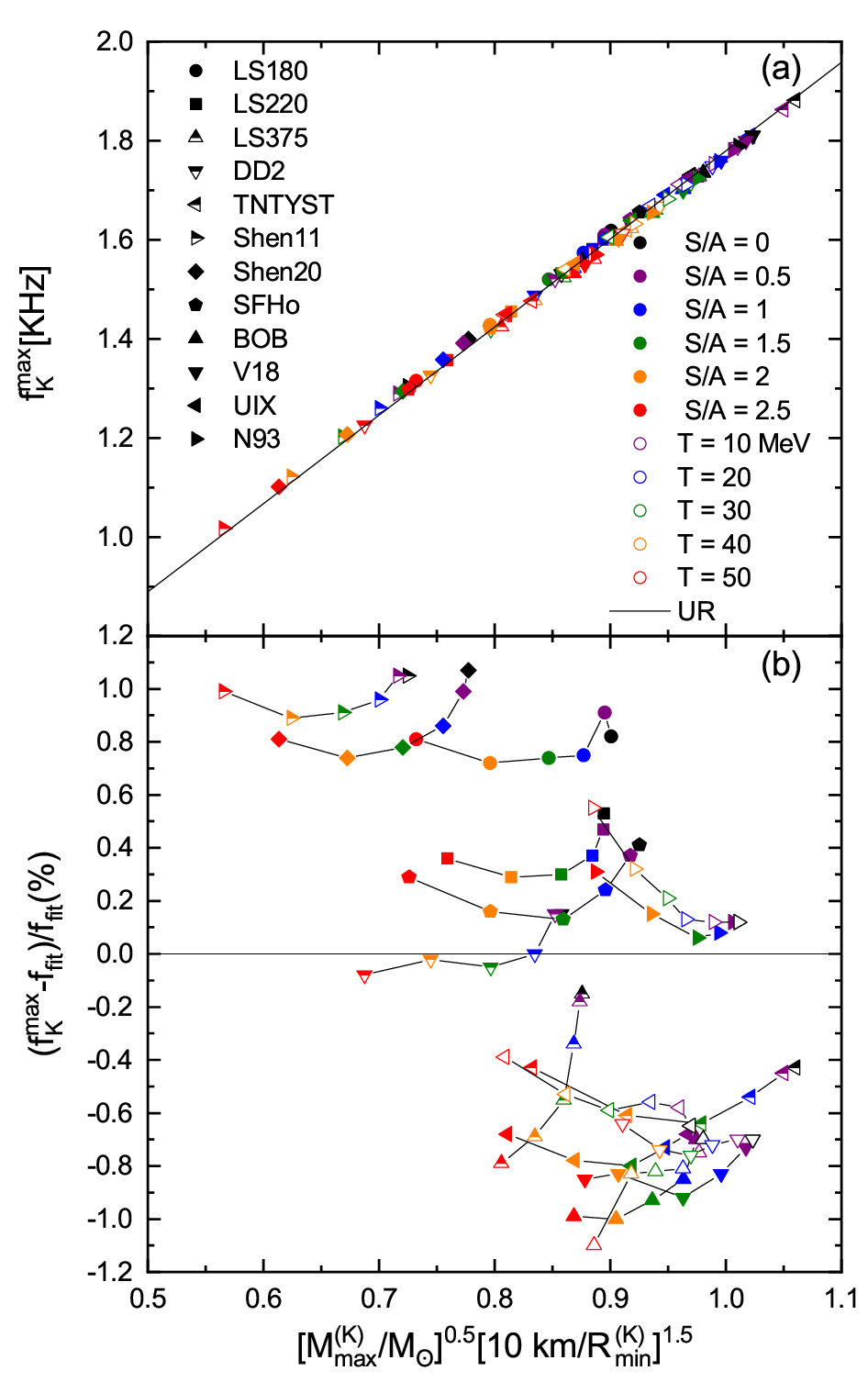}
\vspace{-0mm}
\caption{
(a) Maximum Kepler frequency $\fkm$
as a function of the quantity
$\big({\mkm}/{\!\ms}\big)^{1/2} \big({10\km}/{\rkm}\big)^{3/2}$,
Eq.~(\ref{e:ffita}),
for the rotating PNS maximum-mass configurations
for different EOSs and thermal conditions.
The solid line shows the UR Eq.~(\ref{e:ffitb}).
(b)
The relative deviation between the maximum Kepler frequency $\fkm$
and the UR Eq.~(\ref{e:ffitb}).
}
\hj{}
\label{f:ffit}
\end{figure}%...................................................................

\begin{figure*}%................................................................
\vspace{-2mm}\hspace*{-2mm}
\includegraphics[width=1\textwidth]{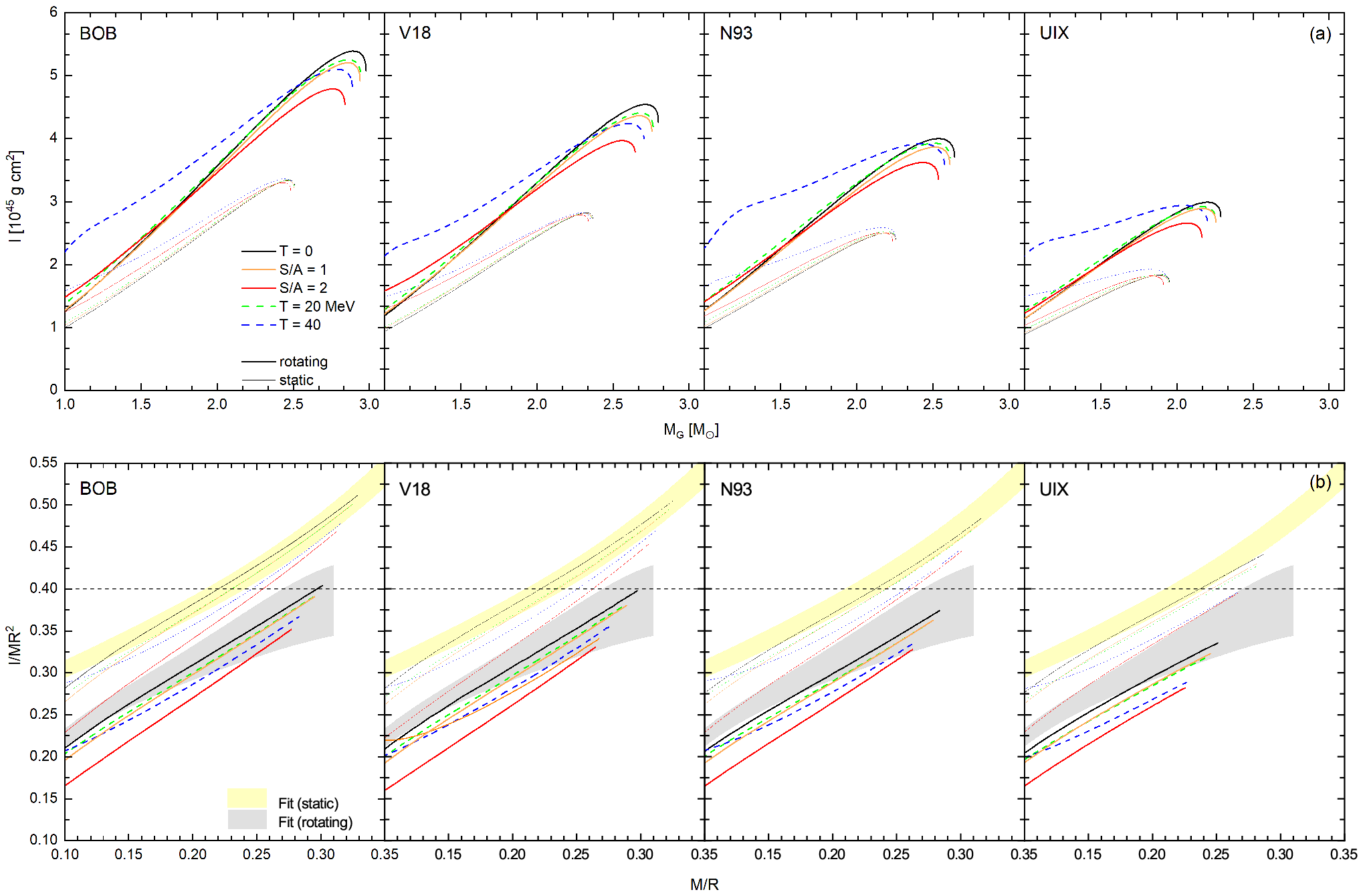}
\vspace{-1mm}
\caption{
(a) MOI $I$ for static and Kepler configurations
as a function of gravitational mass
for different EOSs and thermal conditions.
(b) Dimensionless MOI $I/{MR^2}$
as a function of compactness $\beta=M/R$.
The dashed horizontal line is the constant-density limit.
The yellow and grey bands are the URs of \cite{Wei19,Koliogiannis20}
for cold NSs.}
\label{f:I}
\end{figure*}%..................................................................

\begin{figure}%................................................................
\vspace{-1mm}\hspace*{-2mm}
\includegraphics[width=0.49\textwidth]{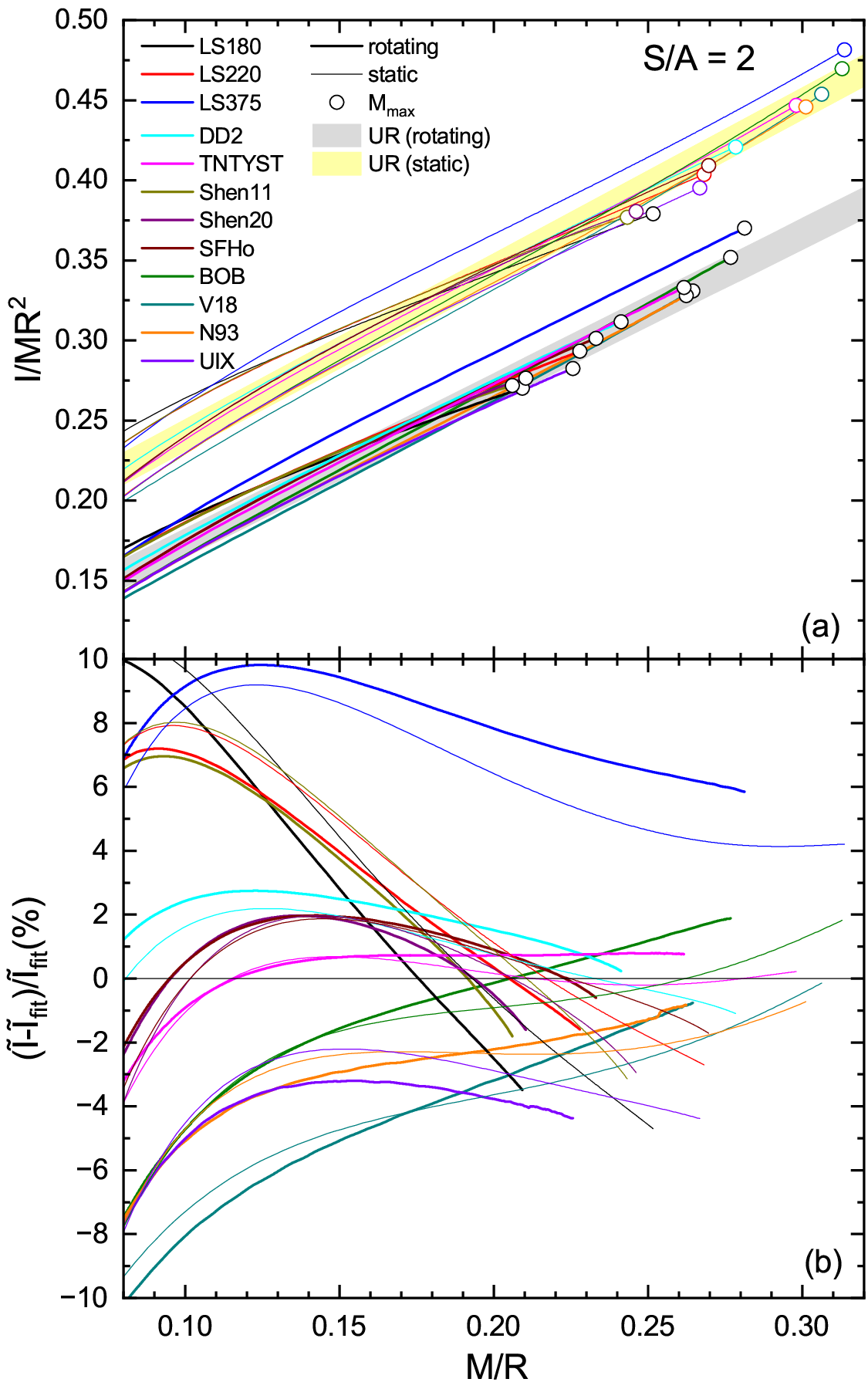}
\vspace{-4mm}
\caption{
(a) $I/MR^2$ as a function of $M/R$ at $S/A = 2$
for the different EOSs shown in Fig.~\ref{f:ffit}.
The yellow and grey bands show the URs Eqs.~(\ref{e:fitIS},\ref{e:fitIK}).
(b) The relative deviation between the calculated $I/MR^2$
and the URs.
}
\label{f:sa2}
\end{figure}%..................................................................

\begin{figure*}%................................................................
\vspace{2mm}\hspace*{-2mm}
\includegraphics[width=1\textwidth]{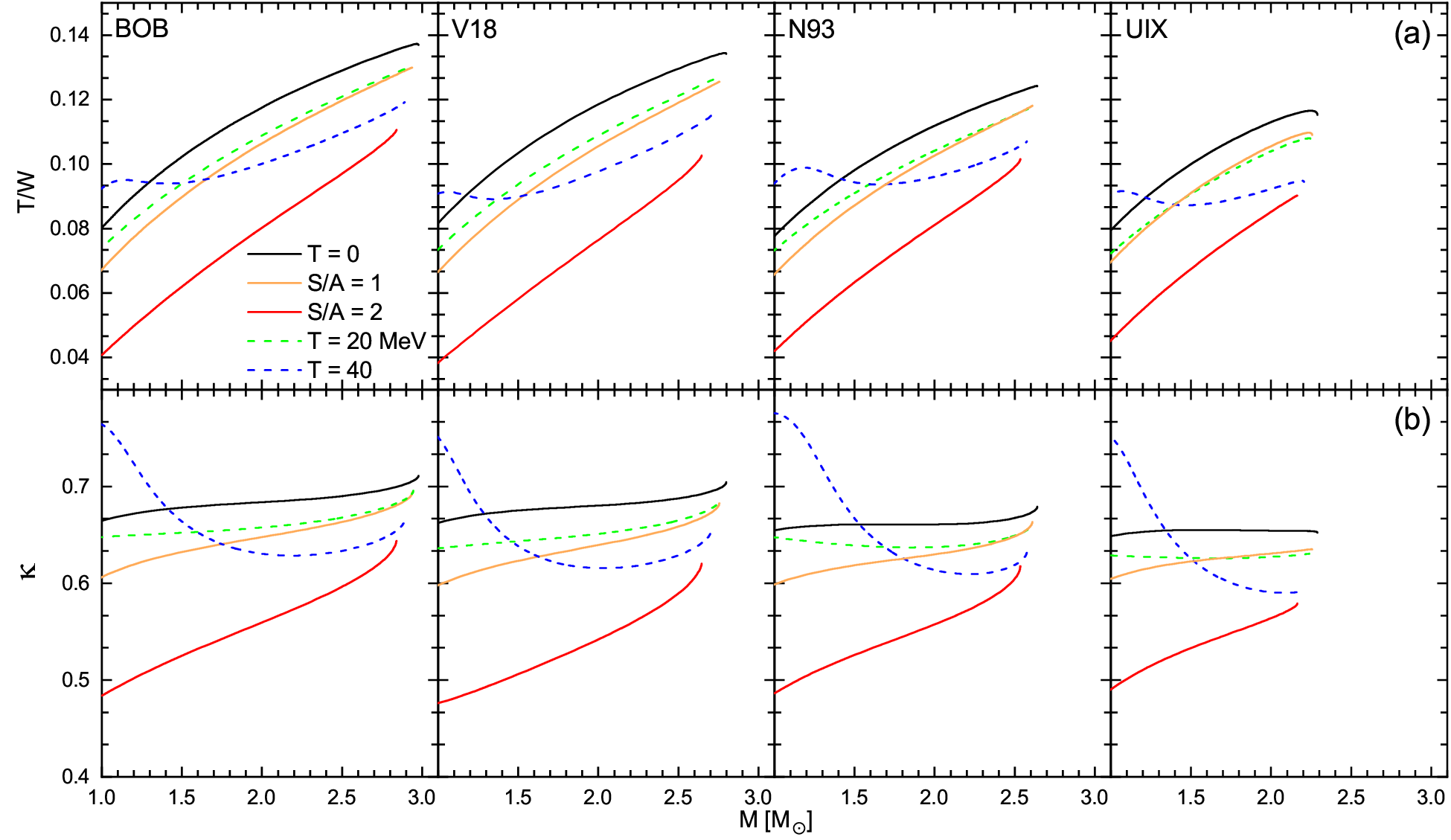}
\vspace{-1mm}
\caption{
(a) The ratio $T/W$
of rotational kinetic energy to the gravitational potential energy, and
(b) Kerr parameter $\kappa=J/M^2$,
at the Kepler frequency
as a function of gravitational mass for different EOSs and thermal conditions.
}
\label{f:twk}
\end{figure*}%..................................................................

%-------------------------------------------------------------------------------
\subsection{Kepler frequency}

The Kepler frequency is an important property of rotating NSs.
In Fig.~\ref{f:fk} it is displayed
as a function of the gravitational mass of rapidly rotating NSs
for different EOSs and thermal conditions.
The current observational lower limit 1.122 kHz from XTE J1739-285
\cite{Kaaret07} is also shown,
and lies well below the theoretically predicted $\fkm$
for the rotating maximum-mass configurations.
As discussed in the previous section,
the Kepler frequency at a given gravitational mass,
and therefore also $\fkm$,
decrease with the temperature.

Recently \cite{Koliogiannis20},
an empirical formula for the absolute maximum frequency
of cold rotating NSs for a given EOS was introduced,
relating it to the mass and radius of those Kepler configurations,
\bal
 \fkm(T=0) &\approx
% C_\text{rot}
 1.78\;\text{kHz}
 \left( \frac{\mkm}{\ms} \right)^{1/2}
 \left( \frac{10\km}{\rkm} \right)^{3/2}
% \equiv C x_{\rm{max}}^{\rm{rot}} \:,
\label{e:ffita}
\\
 &\approx 0.155 \Big(\mkm/{\rkm}^3\Big)^{1/2}
\\
 &\equiv f_\text{fit}(\mkm,\rkm) \:,
\label{e:ffitb}
\eal
%with $C_\text{rot} = 1.78\;$kHz,
which holds to within 1\% relative error for 23 different EOSs.
In Fig.~\ref{f:ffit} we examine the validity of this fit at finite temperature
for the BHF EOSs along with several others,
including Shen11, Shen20, HS(DD2), TNTYST, SFHx, LS180, LS220, and LS375,
which are publicly available on COMPOSE \cite{compose}.
As shown in Fig.~\ref{f:ffit}(b),
the relative deviation between the maximum Kepler frequency $\fkm$
and the UR Eq.~(\ref{e:ffitb})
is less than about 1\%.
Indeed the UR is still valid to the same accuracy as for cold stars.
Also \cite{Khadkikar21} studied a set of relations
between global parameters of hot and rapidly rotating NSs,
and it was shown that Eq.~(\ref{e:ffita})
remains valid at constant $S/A$
and constant electron lepton fraction $Y_e$.

We note that recent studies
\cite{Raduta20,Largani22,Tsiopelas26}
have shown that certain URs,
such as the $I$-Love-$Q$ relations,
are sensitive to thermodynamic conditions
as they depend heavily on the stellar density profile.
However, the UR Eq.~(\ref{e:ffita}) explored in this work,
which relates the global maximum properties of rotating configurations,
belongs to a different class of `limiting-property' relations.
This relation is primarily governed
by the bulk gravitational potential ($M/R^3$)
and the balance between gravity and centrifugal force at the stellar surface.
Consequently,
it remains remarkably robust against thermal effects and EOS variations.

%-------------------------------------------------------------------------------
\subsection{Moment of inertia}

The MOI is an important physical quantity
to describe the rotation characteristics of NSs
and is roughly proportional to the product of the NS gravitational mass
and squared radius.
For rigid rotation it can be defined as
$I = J/\Omega$,
where $J$ is the angular momentum
and $\Omega \equiv 2\pi f$ is the angular frequency
measured by a distant observer (pulsar frequency).
As reported in \cite{Lyne04,Lattimer05,Kramer09},
the MOI can be estimated by timing observations
of binary radio pulsar systems,
which provides crucial insights into the EOS at high density.

Figure~\ref{f:I}(a) shows the MOI
of static and Kepler configurations
as a function of gravitational mass
for the BHF EOSs with different thermal conditions.
It increases almost linearly with the NS mass for all EOSs,
corresponding to nearly constant radii over a large mass range,
see Fig.~\ref{f:mr}. %(b).
The stiffness of the EOS clearly influences the MOI,
the stiffest BOB EOS (large radii) predicts a larger MOI for a given mass.
For temperatures $T\lesssim20\mev$ and not too close to the maximum mass,
thermal effects are small,
but they become notable for $T\gtrsim20\mev$ and $S/A=1,2$,
due to the increase of stellar radii,
as shown in Fig.~\ref{f:mr}. %(b).

Fig.~\ref{f:I}(b) shows the dimensionless MOI
$\tilde{I} \equiv I/(MR^2)$
as function of the compactness $\beta \equiv M/R$.
Nearly linear relations are observed for all EOSs and thermal conditions.
Massive stars have `harder' density profiles
and are thus closer to the constant-density limit $I/(MR^2)=2/5$.
Higher temperatures reduce the MOI for fixed compactness.
The yellow and gray bands represent the polynomial fits
given for cold static and rotating NSs
in \cite{Wei19,Koliogiannis20}, respectively.
As can be seen the cold configurations (solid black curves) for all EOSs
fulfill these URs,
but finite temperature reduces the ratio $I/MR^2$,
out of the $T=0$ fit range.

Following \cite{Raduta20} we therefore attempt to fit the results
for fixed thermal conditions.
As an example,
Fig.~\ref{f:sa2}(a) shows the dimensionless MOI $I/MR^2$ as a
function of the compactness $\beta=M/R$
for the different EOSs shown in Fig.~\ref{f:ffit},
at fixed $S/A$ = 2.
All curves increase almost linearly with $M/R$
and lie inside a `universality band'.
Following the linear fit method for cold and static NSs
reported in \cite{Wei19},
our results are
\bal
 I/MR^2 &= 0.137 + 1.036\beta \pm 0.01 \quad  \text{(Static)} \:,
\label{e:fitIS}
\\
 I/MR^2 &= 0.077 + 0.969\beta \pm 0.01 \quad  \text{(Kepler)} \:,
\label{e:fitIK}
\eal
which are shown as yellow and grey bands.
Panel (b) shows that the relative deviations from these linear URs
are less than 10\% for all EOSs.
Similar conclusions have been drawn by Ref.~\cite{Raduta20}
for hot NS matter with $S/A$ = 2 and $Y_e$ = 0.2,
where polynomial fits of higher order were adopted,
but the combination with untrapped matter was not studied there.

%-------------------------------------------------------------------------------
\subsection{The ratio $T/W$}

The ratio of the rotational kinetic energy at the Kepler frequency, $T$,
and the gravitational potential energy, $W$,
is an important parameter used to characterize the stability of rotating NSs.
It has been indicated that once $T/W$ exceeds a critical value of about 0.27,
a rigidly-rotating NS becomes dynamically unstable,
potentially causing it to collapse or erupt material
\cite{Tohline85,Rather21b}.
In Fig.~\ref{f:twk}(a) we show $T/W$ as a function of gravitational mass
for different EOSs.
It increases due to the increase of the Kepler frequency with mass.
The maximum values of less than 0.14 are far from the instability limit.
Finite temperature reduces $f_K$ and therefore also $T/W$.
It is important to note that this analysis assumes rigid rotation of the NS,
whereas differential rotation would alter the critical $T/W$ value
\cite{Pickett96,Shibata00,Saijo06,Saijo08,Fu11}.

%-------------------------------------------------------------------------------
\subsection{The Kerr parameter}

The so-called Kerr black holes are realized by the Kerr space-time, i.e.,
general stationary vacuum solutions of the Einstein field equations,
which are completely characterized by solely
angular momentum $J$ and gravitational mass $M$
of rotating black holes \cite{Lo11,Maselli13}.
The Kerr parameter is defined as
$\kappa \equiv J/M^2 = \Omega I/M^2$.
Maintaining $\kappa<1$ ensures a stable Kerr black hole,
whereas $\kappa>1$ will lead to a naked singularity without a horizon.
Due to the conservation of angular momentum during the gravitational collapse
of a massive rotating NS,
the resulting black hole may have a mass and angular momentum
similar to those of the NS.
For this reason, the study of the Kerr parameter
provides insight into the NS rotational properties,
stability, and final fate during gravitational collapse
\cite{Lo11,Koliogiannis20,Rather21}.

Fig.~\ref{f:twk}(b) displays the Kerr parameter
as a function of gravitational mass
of rapidly rotating NSs at the Kepler frequency
for different EOSs.
The values for cold configurations across all EOSs
range from approximately 0.65 to 0.7,
which is compatible with the results in \cite{Lo11}.
Thermal effects cause a slight reduction,
mainly due to the decrease of the Kepler frequency at low mass,
and the MOI at high mass,
see Figs.~\ref{f:fk} and \ref{f:I}.

%-------------------------------------------------------------------------------
\section{Summary}
\label{s:conc}

We investigated thermal effects
on the properties of rapidly-rotating NSs up to the Kepler frequency
by employing the BHF approach at finite temperature
with four different $NN$ potentials and compatible nuclear three-body forces.
We found that the maximum gravitational mass
decreases with increasing temperature,
due to an earlier onset of mass shedding
correlated with an expansion of the radius at the Kepler frequency.
We then examined the effects on other stellar properties,
including the Kepler frequency,
the MOI,
the ratio of the rotational kinetic energy and gravitational potential energy,
and the Kerr parameter.

The Kepler frequency at a given gravitational mass
decreases with increasing temperature for all EOSs.
The universal linear relation
established for cold NSs
between the maximum Kepler frequency and the quantity
$\big({\mkm})^{1/2}/({\rkm})^{3/2}$
at the maximum-mass configuration
is confirmed also at finite temperature.

The MOI increases with increasing gravitational NS mass for all EOSs,
but decreases before reaching the maximum mass,
in accordance with the radius.
For a fixed small/large mass,
thermal effects increase/decrease the MOI,
again in accordance with the thermal change of the radius.
The nearly linear $T=0$ URs between $I/(MR^2)$ and $M/R$
are violated at finite temperature.
However, the relations become EOS-independent under the isentropic condition,
consistent with the conclusion in Ref.~\cite{Raduta20}.

The ratio of $T/W$ for rotating NSs remains well below
the instability onset 0.27 across the entire temperature range.
The predicted maximum Kerr parameter values for all cold EOSs
are approximately 0.7.
Thermal effects on the Kerr parameter for mid-mass rotating NSs
at high temperatures are significant due to the pronounced expansion.

The calculations were performed under the assumption of
idealized isothermal or isentropic temperature profiles,
and in particular absence of neutrino trapping,
which might not be adequate for realistic astrophysical scenarios.
We plan to employ various microscopic EOSs in detailed simulations
to compare with future observations of merger events.

\section*{Acknowledgments}

This work is sponsored by
the National Key R$\&$D Program of China
No.~2022YFA1602303,
the National Natural Science Foundation of China under Grant
No.~122-
05260,
the Shandong Provincial Natural Science Foundation, China
ZR2025QC1487
and the Qingdao Natural Science Foundation 25-1-1-4-zyyd-jch.

%-------------------------------------------------------------------------------
\def\epja{Euro. Phys. J. A}
\def\aap{Astron. Astrophys.}
\def\apj{Astrophys. J.}
\def\apjl{Astrophys. J. Lett.}
\def\jcap{Journal of Cosmology and Astroparticle Physics}
\def\jcap{JCAP}
\def\mnras{Mon. Not. R. Astron. Soc.}
\def\nphysa{Nucl. Phys. A}
\def\physrep{Phys. Rep.}
\def\plb{Phys. Lett. B}
\def\ppnp{Prog. Part. Nucl. Phys.}
\def\prc{Phys. Rev. C}
\def\prd{Phys. Rev. D}
\def\prl{Phys. Rev. Lett.}
\def\url#1{#1}
\bibliographystyle{epj}
\bibliography{rpns}

\end{document}